\documentstyle[prl,aps,epsf]{revtex}

\begin{document}
\draft
\wideabs{
\title{Multiple Andreev reflections and enhanced shot noise 
in diffusive SNS junctions}
\author{E.V. Bezuglyi$^{a,b}$, E.N. Bratus'$^{a,b}$, V.S. Shumeiko$^{b}$, and
G. Wendin$^b$}
\address{$^a$B.Verkin Institute for Low temperature Physics and
Engineering, Kharkov, Ukraine}
\address{$^b$Chalmers University of Technology and G\"oteborg University,
41296 G\"oteborg, Sweden}

\maketitle

\begin{abstract}
We study the dc conductance and current fluctuations in diffusive
voltage biased SNS junctions with a tunnel barrier inside the mesoscopic
normal region. We find that at subgap voltages, $eV<2\Delta/n$, the
current associated with the chain of $n$ Andreev reflections is mapped
onto the quasiparticle flow through a structure of $n+1$ voltage biased
barriers connected by diffusive conductors. As a result, the
current-voltage characteristic of a long SNINS structure obeys Ohm's
law, in spite of the complex multiparticle transport process.  At the
same time, nonequilibrium heating of subgap electrons produces giant
shot noise with pronounced subharmonic gap structure which corresponds
to stepwise growth of the effective transferred charge.
At $eV\rightarrow 0$, the shot noise approaches the magnitude of
the Johnson-Nyquist noise with the effective temperature
$T^\star=\Delta/3$, and the effective charge increases as $(e/3)(1 +
2\Delta/eV)$, with the universal ``one third suppression'' factor. We
analyse the role of inelastic scattering and present a criterion of
strong nonequilibrium.
\end{abstract}
}
\narrowtext

Current transport through mesoscopic resistive elements (tunnel
barriers and disordered normal conductors) attached to superconductors
is a subject of permanent interest and intensive experimental studies.
The investigations of superconducting junctions are primarily focused
on the complex nonlinear behavior of current-voltage characteristics, which
exhibit subharmonic gap structure, zero bias anomaly, etc. In this
Letter, we will discuss an elementary mesoscopic superconducting
structure where the current shot noise manifests anomalous transport
properties while the average current shows perfect Ohmic behavior.

The circuit under discussion consists of a low-trans\-mis\-sion tunnel
junction (or point contact) connected to voltage biased superconducting
reservoirs via diffusive normal leads (Fig.\ 1a), like e.g. a
short-gate Josephson field effect transistor \cite{Transistor}. In a NIN
structure connected to normal reservoirs, the average current obeys
Ohm's law and the current fluctuations show full Poissonian shot noise,
$S=2eI$, if the tunnel resistance $R$ dominates over the resistance of
the normal leads \cite{Jong}. The effect of the superconducting reservoirs, 
which has recently attracted much attention, is to modify the density of
states and to create a gap $E_g$ in the electron spectrum of the normal
leads \cite{Gap}. This proximity effect provides dc Josephson current
flow and, simultaneously, blocks the single-particle tunneling at
applied voltages $eV < 2E_g$. At these subgap voltages, the current is
due to multiparticle tunneling (MPT) \cite{SW}. The MPT regime is
manifested by the stepwise decrease of the current with decreasing
applied voltage (subharmonic gap structure at $eV=2E_g/n$), which
provides exponential decay of the current \cite{Bratus}. At the same
time, the current shot noise undergoes enhancement due to the growth of
the elementary tunneling charge $ne$ \cite{Imam}. MPT has been
extensively studied theoretically in quantum point contacts
\cite{Bratus,Ave1}, and in short diffusive constrictions \cite{Averin}
with a wide proximity gap of the order of the energy gap $\Delta$ in
the superconducting reservoirs. Both the subharmonic gap structure and
the enhanced shot noise have been observed experimentally
\cite{Jan,Expnoise}.

A distinctly different transport regime occurs in long diffusive SNS
junctions with a small proximity gap of the order of the Thouless energy,
$E_g\sim E_{Th}\ll\Delta$. In this case, the Josephson current is
suppressed, and single-particle tunneling dominates at virtually all
applied voltages. However, when the inelastic mean free path exceeds the
distance between the superconducting reservoirs, current transport
at subgap voltages $eV<2\Delta$ is still non-trivial: the tunneling
electrons must undergo multiple Andreev reflections (MAR) before they
may enter the reservoirs \cite{KBT}. We will show that in long SNS
structures with opaque tunnel barriers, the current-voltage
characteristic is perfectly linear and structureless, while the current
shot noise is greatly enhanced and reveals subharmonic gap structure
(kinks) at voltages $eV=2\Delta/n$ (incoherent MAR regime).

The origin of the linear current-voltage dependence and the significant
deviation of tunnel shot noise from the Poisson law can be
qualitatively explained in the following way. In order to overcome the
energy gap at low voltages ($eV\ll 2\Delta$) the electron has to undergo a
large number ($M\approx 2\Delta/eV$) of Andreev reflections, gaining the
energy $eV$ in each passage of the tunnel barrier (Fig.\ 1b). Thus,
MAR tunneling in real space is associated with probability current flow
along the energy axis through a structure of $M+1$ tunnel barriers with
the total effective resistance $R_M = (M+1)R$. Since only electrons
incoming within the energy layer $eV$ below the gap $2\Delta$
participate in MAR transport, the total probability current is $I_p =
V/R_M$. However, each pair of consequtive Andreev reflections transfers
the charge $2e$ through the junction, and the real current $I$ is
therefore $M+1$ times greater than the probability current: $I=(M+1)I_p
= V/R$. The current flow in energy space generates shot noise $S_p$
which is related to the probability current as $S_p=(2/3)eI_p$ in the
limit $M\rightarrow\infty$. The arguments for the
1/3 suppression of the Poissonian noise in multibarrier tunnel
structures are similar to the ones presented in Ref.\ \cite{Jong}. 
Since the noise spectral density
is given by the current-current correlation function, the real shot
noise $S$ is $(M+1)^2$ times greater than $S_p$, i.e. approaches a
constant value $S=(4/R)(\Delta/3)$. This coincides with the exact result,
Eqs.\ (\ref{ShotNoise1}), (\ref{ChargeV}) below, in the limit
$eV\rightarrow0$.

\begin{figure}[tb]
\epsfxsize=7.5cm\centerline{\epsffile{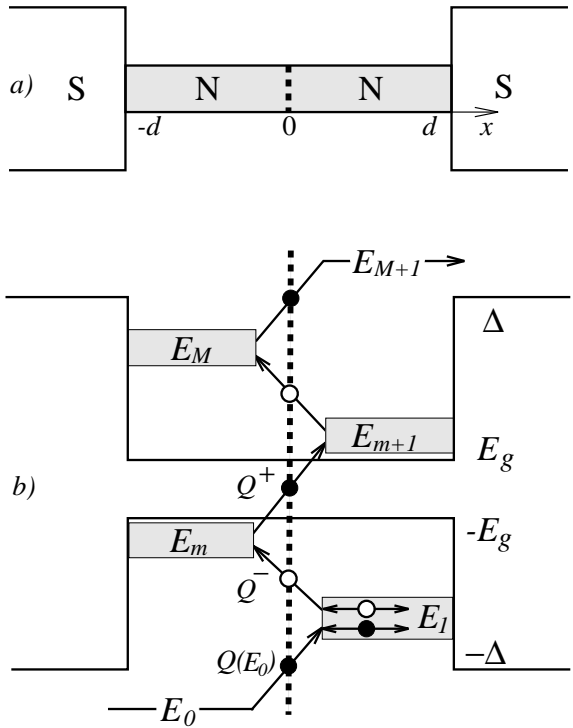}}
\vspace{0.1in}
\caption{a) Diffusive mesoscopic junction with a tunnel barrier at
$x=0$ (dashed line). b) Schematic picture of incoherent MAR. An
equilibrium electron incoming from the superconducting reservoir
consequently tunnels through $M$ Andreev side bands and creates
elementary probability current $Q$ flowing along the energy axis
between the equilibrium regions outside the gap. Each crossing of
the barrier is associated with the transfer of the elementary charge
$e$, and the electric current is $eQ(M+1)$.}
\end{figure}

For a quantitative treatment of incoherent MAR, we consider the
diffusive kinetic equation \cite{LO1} for the 4$\times$4 supermatrix
Keldysh-Green's function $\check G(x,t_1,t_2)$ 
\begin{equation} \label{Keldysh}
i\nabla \check{J} =  [\check{H}, \check G], \quad 
\check G^2 = \check{1}, \quad
\check{J} = {\cal D}\check G\nabla\check G,
\end{equation}
where ${\cal D}$ is the diffusion constant. We apply
Eqs.\ (\ref{Keldysh}) to the electrons in the normal leads and match
the Green's functions at the tunnel barrier ($x=\pm 0$) of low transparency
($\Gamma\ll 1$) using the boundary condition \cite{KL}
\begin{equation} \label{KeldyshBoundary} 
\check{J}(-0)\! = \!\check{J}(+0)\! =
\!(j/2)[\check G(-0)\!,\check G(+0)], \; j\! = \!(1/2)v_F\Gamma.
\end{equation}
Equation (\ref{KeldyshBoundary}) represents the conservation law for the
supermatrix current $\check{J}$ and connects it with the
voltage-induced imbalance of elementary probability currents $j$ of
tunneling electrons. Since we have assumed that the barrier resistance
$R$ dominates over the resistance $R_d$ of the leads ($R\gg R_d$), the
voltage drop at the barrier determines the time-dependent phase
difference between the reservoirs by virtue of the Josephson relation.
A gauge transformation allows us to remove the time dependence in the
Hamiltonian in Eq.\ (\ref{Keldysh}) and cancel the electric potential.
Simultaneously, a periodic time dependence appears in the boundary
condition of Eq.\ (\ref{KeldyshBoundary}), which implies that the
Green's function $\check G(x,t_1,t_2)$ consists of a set of harmonics
$\check G(x,E_n,E_m)$, $E_n=E+neV$. The problem is solvable in the
limit of weak Josephson coupling, $\Gamma\ll 1$ and/or $E_{Th} \ll
\Delta$, when off-diagonal Green's function harmonics can be neglected
and the diagonal harmonics ($n=m$) satisfy the static equations,
similar to the case of zero applied voltage.

The dc current $I$ can be expressed, to first order in $\Gamma$,
through the quasiparticle distribution function $f(E)$ defined by the
following representation of the Keldysh function: $\hat g^K = \hat g^R
\hat f\! - \!\hat f\hat g^A$, $\hat{f} = \hat{1}(1\!-\!f) + \sigma_z f_z$,
where $\hat g^{R,A}$ are the retarded (advanced) Green's functions,
giving
\begin{equation} \label{TunnelModel}
I=\!{1\over 2eR}\!\!\int^{+\infty}_{-\infty}\!\!\! dE N(E) N(E\!+\!eV)
[f(E)\!-\!f(E\!+\!eV)].
\end{equation}
The density of states, $N(E)=(1/4)\mbox{Tr}\sigma_z(\hat g^R - \hat
g^A)$, is calculated for a nontransparent barrier and 
normalized by the normal-electron density of states; all values are
taken at the interface, $x=0$.  Eq.\ (\ref{TunnelModel}) has the same
form as the conventional equation of the tunnel model \cite{Cohen},
with the nonequilibrium distribution function $f(E,x)$ obeying a
kinetic equation following from Eq.\ (\ref{Keldysh}),

\begin{equation} \label{KineticEq}
{\partial\over\partial x}D(E,x){\partial f\over\partial x}\! =
\!{N(E,x)\over\tau_\varepsilon} [f(E,x)\! - \!f_0(E)], 
\end{equation}
$D(E,x) = ({\cal D}/ 4)\mbox{Tr}(1 - \hat g^R \hat g^A)$. The inelastic
scattering term in Eq.\ (\ref{KineticEq}), describing relaxation to
equilibrium population $f_0(E)=2n_F(E)$, is written for simplicity in
the relaxation time approximation. The boundary condition for the
function $f$ at the tunnel interface obtained from
Eq.\ (\ref{KeldyshBoundary}) reads
\begin{equation} \label{KineticBoundary}
\left. -D(E,x)(\partial f/\partial x)\right|_0 = Q^+(E) - Q^-(E), 
\end{equation}
\begin{equation} \label{SpectralCurrent}
Q^\pm(E)\! = \!\pm(j/ 2) N(E)N(E\!\pm\! eV)[f(E)\!-\!f(E\!\pm \!eV)].
\end{equation}

The quantities $Q^\pm$, which also determine the current in Eq.\
(\ref{TunnelModel}), can be interpreted as spectral quasiparticle
currents, i.e. probability currents flowing upwards in energy space
(Fig.\ 1): the current $Q^+$ exits from energy $E$ towards energy
$E+eV$, while the current $Q^-$ arrives at energy $E$ from energy
$E-eV$. Along this line of reasoning, the boundary condition,
Eq.\ (\ref{KineticBoundary}), represents the detailed balance between
the spectral quasiparticle current and the leakage current [the term on
the left hand side of Eq.\ (\ref{KineticBoundary})] due to either
inelastic relaxation or escape into the reservoirs.

Let us consider the limit of infinitely large inelastic relaxation
time \cite{note3}.  In this case, the leakage current is
spatially homogeneous according to Eq.\ (\ref{KineticEq}). Within the energy gap,
$|E|<\Delta$, the diffusion coefficient $D(E,x)$ turns to zero in the
superconducting reservoirs, and therefore the leakage current is
blocked, indicating complete Andreev reflection. Thus, the spectral
current $Q^\pm$ is conserved within the superconducting gap:
$Q^+=Q^-$.  This equation provides recurrence relations for the
nonequilibrium distribution functions $f(E_n)$ in different side bands
associated with MAR. The boundary conditions are established by
the requirement of equilibrium outside the gap, $f(E_n)=2n_F(E_n)$,
$|E_n|>\Delta$.  Indeed, the reservoirs maintain the equilibrium at the
NS boundaries, $f(E,\pm d) = 2n_F(E)$; on the other hand, the gradient
of the distribution function given by Eq.\ (\ref{KineticBoundary}) is
small, $\sim \!R_d/R\!\ll\! 1$, and may be neglected. We note that the
latter condition is equivalent to a small ratio between the diffusion
time through the normal lead, $d^2/{\cal D} =E_{Th}^{-1}$, and the
inverse tunneling rate, $(\Gamma v_F/d)^{-1}$, i.e. $\Gamma v_F/d\ll
E_{Th}\ll\Delta$.

The physical picture of MAR in diffusive SNINS systems is illustrated
in Fig.\ 1b. The equilibrium electron-like quasiparticles incoming from
the left electrode with energy $-\Delta-eV < E_0 < -\Delta$ create a
probability current $Q^\uparrow(E_0)=j n_F(E_0)N(E_0)N(E_0+eV)$ across
the tunnel junction into the subgap region. Due to low transmissivity
of the barrier and fast electron diffusion through the normal leads,
the particle undergoes many Andreev reflections from the superconductor
before the next tunneling event will occur and, therefore, the electron
and hole states at energy $E_1=E_0+eV$ are occupied with equal
probability $(1/2)f(E_1)$ \cite{note2}. Thus, the population $f(E_1)$
produces both the current of holes $Q^\uparrow(E_1)=(j/2)f(E_1)N(E_1)
N(E_2)$ moving upwards along the energy axis to the next side band, and
the counter current of electrons $Q^\downarrow(E_1)=
(j/2)f(E_1)N(E_0)N(E_1)$ down to the initial state, determining the
net probability current $Q(E_0)=Q^\uparrow(E_0)-Q^\downarrow(E_1)$, and
so on. As a result, the electron tunneling in real space is associated
with the flow of spectral current $Q^+(E_m) = Q^-(E_m) = Q(E_0)$
through $M+1$ tunnel barriers connected in series by a number $M(E_0) =
[(\Delta-E_0)/ eV]$ of Andreev side bands ($[x]$ denotes the integer
part of $x$). In this transport problem in energy space, there is an effective
bias voltage $(M+1)eV$ drop between the reservoirs represented by the
spectral regions outside the energy gap, $|E|>\Delta$, and it is
equally distributed among the tunnel barriers. Therefore, the
distribution function has a steplike form,
\begin{equation} \label{Population}
f(E_m)\!=\!2\!\left[\!\left[n_F(\!E_{M+1})\!-\!n_F(\!E_0)\right]\!
{Z_m\over Z_{M+1}}\!+\!n_F(\!E_0)\right],
\end{equation}
\begin{equation} \label{Z}
Z_m(E_0)=\sum_{k=0}^{m-1}N^{-1}(E_k)N^{-1}(E_{k+1}), \; Z_0 = 0.
\end{equation}

The tunnel current in Eq.\ (\ref{TunnelModel}) is determined by the
spectral current $Q^+$. At low temperature, the equilibrium spectral
current at $|E|>\Delta$ is exponentially small, and the main
contribution to the total current comes from the nonequilibrium subgap
region. Dividing it into pieces of length $eV$ and taking into account
spectral current conservation, one finds from
Eqs.\ (\ref{TunnelModel}), (\ref{Population})
\begin{equation} \label{SubgapCurrent}
I(V)\!=\!\!{1\over eR}\!\int^{-\!\Delta}_{-\!\Delta-eV}\!\!\!\!\! dE_0
{M+1\over Z_{M+1}} \!\left(n_F(E_0)\!-\!n_F(E_{M+1})\right).
\end{equation}

Equation (\ref{SubgapCurrent}) describes the single-particle current in
a tunnel junction of arbitrary length.  If some MAR chain contains a
side band $E_n$ within the proximity-induced gap $2E_g$, the
corresponding density of states $N(E_n)$ is zero, and the spectral
current associated with this chain is blocked. At $eV<2E_g$, any MAR
chain has at least one side band within the gap, and the total single
particle current in Eq.\ (\ref{SubgapCurrent}) vanishes. In the limit
of a long junction, the proximity gap closes and the local density of
states becomes constant, $N(E)=1$. In this case, the current in
Eq.\ (\ref{SubgapCurrent}) shows Ohmic behaviour, $I=V/R$, with the same
resistance $R$ as in the absence of superconducting ``mirrors''.

Let us turn to calculation of the tunnel current shot noise power
$S(V)$. A general quantum equation for the shot noise in superconducting
junctions has been derived in \cite{LO2}. Assuming the asymptotic limit
of highly resistive tunnel barrier and the long-junction approximation 
we write it on the form 
\begin{equation} \label{ShotNoise}
S(V) \!=\!\!\!\int^{+\infty}_{-\infty}\!\!{dE\over R}
(f(E)\!+\!f(E\!\!+\!eV)\!-\!\!f(E)f(E\!\!+\!eV)). 
\end{equation}
Taking into account the distribution function in Eq.\ (\ref{Population}),
the noise power at zero temperature becomes 
\begin{equation} \label{ShotNoise1}
S(V) \!=\!{2\over R}\int^{-\Delta}_{-\Delta-eV}{dE\over3} 
\left(M(E)+1+{2\over M(E)+1}\right).
\end{equation}
At voltages $eV\!>\!2\Delta$ this formula gives conven\-ti\-o\-nal Poissonian
noise $S\!=\!2eI$. At subgap voltages, the noise power undergoes
enhancement: it shows a piecewise linear voltage dependence,
$dS/dV=(2e/3R)(1 + 4/([2\Delta/eV]+2))$, with kinks
at the subharmonics of the superconducting gap, $eV_n=2\Delta/n$ (see
Fig.\ 2). At zero voltage, the noise power approaches the constant
value $S(0) = (4/R)(\Delta /3)$, which corresponds to the thermal
Johnson-Nyquist noise with the effective temperature $T^\ast =
\Delta/3$.

The enhancement of the shot noise power can be alternatively 
interpreted as an increase of the effective charge $q(V)=S(V)/2I$ 
with decreasing voltage,
\begin{equation} \label{ChargeV}
{q(V_n) \over e} = {1\over 3}\left(n+1+{2\over n+1}\right) =  1,
{11\over 9}, {22\over15}, \dots
\end{equation}
In the limit $eV \rightarrow 0$ the effective charge increases as
$q(V)/e \approx (1/3)(1+2\Delta/eV)$. This result differs by a factor
1/3 from the value expected from a straightforward MAR argument which
assumes the shot noise to be equal to the Poisson noise enhanced by the
factor $M$. We stress that the 1/3 factor here results from multiple
traversal of the tunnel barrier due to incoherent MAR and has nothing
to do with the diffusive normal leads.

A more detailed analysis with account of inelastic scattering shows
that the current shot noise is suppressed at low voltage when the
lifetime of a quasiparticle within the normal leads becomes comparable
to the inelastic relaxation time ($\alpha \geq 1$, see
Eq.\ (\ref{NoiseInelastic})). Generalized recurrences for the
distribution function in long junctions ($N(E)=1$) then take the form \cite{note1}
\begin{equation} \label{RecurrencesW}
f(E)\!-\!f_0(E)\!=\!W_\varepsilon
\left[f(E\!+\!eV)\!+\!f(E\!-\!eV)\!-\!2f(E)\right].
\end{equation}
The level of nonequilibrium of the subgap electrons is controlled by
the parameter $W_\varepsilon = (R_d/R)(E_{Th}\tau_\varepsilon/ 4)$.
The strong nonequilibrium state discussed above is only
possible at $W_\varepsilon \geq 1$ whereas in the opposite limit,
$W_\varepsilon\ll 1$, the normal leads may always be considered as
reservoirs, and the enhanced noise disappears.

\begin{figure}[tb]
\epsfxsize=7.5cm\centerline{\epsffile{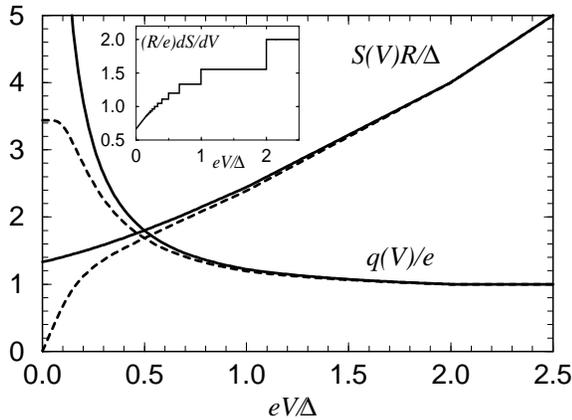}}
\vspace{0.1in}
\caption{Spectral density $S(V)$ of current shot noise and 
effective transferred charge $q(V)=S(V)/2I$ as functions of applied
voltage $V$. In the absence of inelastic collisions (solid lines), the
shot noise power approaches the finite value $S(0)=(4/R)(\Delta/3)$ at $eV
\rightarrow 0$, and the effective charge increases as $q(V) = (e/3) (1
+ 2\Delta/eV)$. The effect of inelastic scattering is represented by
dashed lines for the nonequilibrium parameter $W_\varepsilon = 5$. The
dependence $S(V)$ contains kinks at the gap subharmonics, $eV =
2\Delta/n$, as shown in the inset. }
\end{figure} 

Numerical results for $W_\varepsilon = 5$ are presented in Fig.\ 2 by
dashed curves. The rapid decrease of $S(V)$ at low voltage described by
the following analytical approximation,
\begin{equation} \label{NoiseInelastic}
S(V)\! = \!S(0){3\over \alpha}\!\left( \!\tanh{\alpha\over 2} \!+\!
{\alpha\!-\!\sinh\alpha \over \sinh^2\alpha}\! \right)\!,
\; \alpha \!= \!{\Delta\over eV\!\sqrt{W_\varepsilon}},
\end{equation}
occurs when the length of the MAR chain interrupted by inelastic
scattering, $eV\sqrt{W_\varepsilon}$, becomes less than $2\Delta$.

In conclusion, we have studied subgap tunnel current and current shot
noise in diffusive SNINS structures. We found that in junctions
with normal leads which are much longer than the coherence length but
much shorter than inelastic mean free path, the strongly nonequilibrium
distribution of the subgap electrons created by MAR is manifested in
the shot noise rather than in the tunnel current. While the tunnel
current obeys Ohm's law, the current shot noise is significantly
enhanced and shows subharmonic gap structure.

Support from KVA, NFR, and NUTEK (Sweden), and from NEDO (Japan) is 
gratefully acknowledged.


\begin{references}
\bibitem{note3}
More precisely, $\tau_\varepsilon$ is to be much larger than the
quasiparticle life time within the normal leads, see
Eq.\ (\ref{NoiseInelastic}).
\bibitem{note2}
The equality of electron and hole numbers reflects the absence of an
$f_z$-component of the matrix distribution function $\hat f$, which
assumes the electric field within the normal leads to be negligibly
small.
\bibitem{note1}
This equation is obtained from the kinetic equation
Eq.\ (\ref{KineticEq}) by integrating over the length $d$ of the normal
lead; the assumption of small spatial variations of the distribution
function allows us to replace $f(E,x)$ by its boundary value $f(E)$ in
the collision integral.
\end{references}
\end{document}